\documentclass[aps,pre,twocolumn,groupedaddress,showpacs,floatfix]{revtex4}
\usepackage{graphicx}
\usepackage{amsmath}
\usepackage{amssymb}
\begin{document}
\title{Dependence of kinetic friction on velocity: Master equation approach}
\author{O.M. Braun}
\email[E-mail: ]{obraun.gm@gmail.com}
\homepage[Web: ]{http://www.iop.kiev.ua/~obraun}
\affiliation{Institute of Physics, National Academy of Sciences of Ukraine,
  46 Science Avenue, 03028 Kiev, Ukraine}
\author{M. Peyrard}
\email[E-mail: ]{Michel.Peyrard@ens-lyon.fr}
\affiliation{Laboratoire de Physique de l'Ecole Normale Sup\'{e}rieure de Lyon,
  46 All\'{e}e d'Italie, 69364 Lyon C\'{e}dex 07, France}
\date{\today}
\begin{abstract}

We investigate the velocity dependence of kinetic friction with a
model which makes minimal assumptions on the actual mechanism of
friction so that it can be applied at many scales provided the system
involves multi-contact friction. Using a recently developed master
equation approach we investigate the influence of two concurrent
processes.  First, at a nonzero temperature thermal fluctuations allow
an activated breaking of contacts which are still below the threshold.
As a result, the friction force monotonically increases with velocity.
Second, the aging of contacts leads to a decrease of the friction
force with velocity. Aging effects include two aspects: the delay in
contact formation and aging of a contact itself, i.e., the change of
its characteristics with the duration of stationary contact.  All
these processes are considered simultaneously with the master equation
approach, giving a complete dependence of the kinetic friction force
on the driving velocity and system temperature, provided the interface
parameters are known.

\end{abstract}
\maketitle

\section{Introduction}
\label{intro}

Almost three centuries ago Charles Coulomb (1736-1806) discovered that
kinetic friction does not depend on the sliding velocity~\cite{Dowson1979}.
Later, more careful experiments showed that this law is only
approximately valid~\cite{Persson1998,RM2000,R2000,P2003a,BC2006,BN2006}.
Friction does depend on the sliding velocity, but this dependence is far
from universal: some measurements find an increase when velocity
increases, while others find a decay \cite{HAHSS2007,CRPS2006,Marone}
or even a more complex non-monotonous behavior \cite{Persson1998}.
A logarithmic dependence, often quoted, has been found for two
extreme scales, friction at the tip of an AFM
(see for instance
\cite{HAHSS2007,CRPS2006,Bouhacina,Gnecco,Schirmeisen})
or at the scale
of a fault in the earth crust \cite{Marone}, but it is often only
approximate and observed in a fairly narrow velocity range.
Therefore understanding
the velocity dependence of kinetic friction is still an open
problem, and what makes it difficult is that several phenomena
contribute, the thermal depining of contacts, their aging, and the
delay in contact formation.

\smallskip
In this study we investigate the velocity dependence of friction with
a model that includes these three contributions and makes minimal
assumptions on the actual mechanism of friction so that it can be
applied at many scales provided the system involves multi-contact
friction. Our aim is to elucidate the respective role of these three
contributions to the velocity dependence of friction and to provide
analytical treatments in some limits, or simple numerical approaches
that allow the investigation of a velocity range that may span many
orders of magnitude.

\smallskip
At the most fundamental level multi-contact friction can be described
as resulting a succession of breaking and formation of local contacts
which possess a distribution of breaking thresholds.
This viewpoint was first applied to describe
earthquakes~\cite{BK1967,OFC1992} and then adopted to friction by
Persson~\cite{P1995}.

We recently developed a master equation (ME) approach to describe the
breaking and attachment events~\cite{BP2008,BP2010}. It splits the
analysis in two 
independent parts: (i) the calculation of the friction force, given by
the master equation provided the statistical properties of the
contacts are known, and (ii) the study of the properties of the
contact themselves, which is system dependent. This method is very
general and allows us to calculate the velocity dependence of friction,
which results from the interplay of two concurrent processes.
First, at a nonzero temperature thermal
fluctuations allow an activated breaking of contacts
which are still below their mechanical breaking threshold.
This phenomenon leads to a monotonic increase of the
friction force $F$ with the velocity $v$.
Second, the aging of contacts~\cite{BR2002,FKU2004,BU2010}
leads to a decrease of the friction force with velocity. It
includes two processes:
the delay in contact formation, i.e., time lag between contact
breaking and re-making~\cite{S1963,FKU2004,SW2009,BP2010,BU2010}, and
the aging of a contact itself, i.e., the change of its characteristics
with the time of stationary contact.  To incorporate the latter
effect, the master equation must be completed by an equation for the
evolution of static thresholds.

In earlier studies \cite{BP2008,BP2010} we considered thermal and aging
effects separately, to set up the method. However, to relate the
results to experiments, both contributions must be taken into account
simultaneously. This is the aim of the present paper, which is
organized as follows. Section~\ref{master} is a brief review of the
master equation approach. Section~\ref{temperature} discusses temperature
effects. Whereas our earlier work \cite{BP2010} focussed on
time-dependent phenomena to analyze stick-slip, here we concentrate on
the steady-state case (constant velocity). This allows us to proceed
further and derive explicit expressions for the influence of
temperature. Then Sec.~\ref{aging} introduces the second effect, the
aging of the contacts. It first summarizes the method introduced
earlier and its main results \cite{BP2010}, which only considered the
$T=0$ case, and then studies the combined influence of aging and
temperature fluctuations. Section~\ref{delay} adds the influence of the
delay in contact formation after breaking, to get the full picture,
allowing us to compute the velocity dependence of
friction. Section~\ref{experiments} discusses all those results in the
context of experimental data. The difficulty to apply the theory to
actual experiments is to properly assess the values of the parameters
that enter in the theoretical expressions, and not simply try to fit
experimental curves, which would not be very significant owing to the
number of parameters which are involved. Therefore
Sec.~\ref{experiments} focusses on this assessment.
Finally,
Sec.~\ref{concl} concludes the paper with a discussion of perspectives
for its further development.

\section{Master equation}
\label{master}

The earthquake (EQ) model is the most generic model for friction due
to multiple contacts at an interface.  The sliding interface is
treated as a set of $N_c$ ``contacts'' which deform elastically with
the average rigidity $k$.
The contacts represent, for example, asperities for the interface of
rough surfaces~\cite{GW1966}, or patches of lubricant or its domains
(``solid islands''~\cite{P1993b}) in the case of lubricated friction.
The $i$th contact connects the slider and the bottom substrate through
a spring of elastic constant $k_i$.  When the slider moves, the
position of the contact point changes, and the contact's spring elongates or
shortens, so that the slider experiences the force
$-F = \sum f_i$  from the interface,
where $f_i = k_i x_{i}$ and $x_{i} (t)$ is the spring length.
The contacts are coupled frictionally to the slider.
Namely, as long as the force $|f_i|$ is below a certain threshold $f_{si}$
(corresponding to the onset of 
plastic flow of the entangled asperity,
or to local shear-induced melting of the
boundary lubrication layer),
this contact $i$ moves together with the slider.
When the force exceeds the threshold, the contact breaks,
and then re-attaches again in the unstressed state
after some delay time $\tau$.
Thus with every contact we may associate the threshold value $f_{si}$,
which takes random values from a distribution $\widetilde{P}_c (f)$
having a mean value $f_{s}$. 
The spring constants are related to the threshold forces
by the relationship
$k_i = k \, ( f_{si} /f_{s} )^{1/2}$,
because the value of the static threshold is proportional
to the area $A_i$ of the given contact,
while the transverse rigidity $k_i$ is proportional to contact's size,
$k_i \propto \sqrt{A_i}$.
When a contact is formed again (re-attached to the slider),
new values for its parameters have to be assigned.

Rather than studying the evolution of the EQ model by numerical simulation
it is possible to describe it analytically~\cite{BP2008,BP2010}.
Let $P_c (x)$ be the normalized probability distribution of values
of the thresholds $x_{si}$ at which contacts break;
it is coupled with the distribution of threshold forces by the relationship
$P_c (x) \, dx = \widetilde{P}_c (f) \, df$.
To describe the evolution of the model,
we introduce the distribution $Q(x;X)$ of the stretchings $x_i$
when the bottom of the solid block is at a position $X$.
Let us consider a small displacement $\Delta \! X>0$ of the bottom of
the sliding block.
It induces a variation of the stretching $x_i$ of the contacts which
has the same value  $\Delta \! X$ for all contacts
(here we neglect the elastic deformation of the block).
The displacement $X$ leads to three kinds of changes in the
distribution $Q(x;X)$: first, there is a shift due to the global
increase of the stretching of the asperities;
second, some contacts break because their stretching exceeds the
maximum value that they can withstand;
and third, those broken contacts form again,
at a lower stretching, after a slip at the scale of the asperities,
which locally reduces the tension within the corresponding asperities.
These three contributions can be written as a master equation for $Q(x;X)$:
\begin{equation}
\left[
\frac{ \partial }{ \partial x} +
\frac{ \partial }{ \partial X} +
P(x) \right] Q(x;X)
= R(x) \, \Gamma (X) \,,
\label{Q5a}
\end{equation}
where
$P(x) \, \Delta \! X$ describes the fraction of contacts that break
when the slider position changes from
$X$ to $X + \Delta \! X$.
At zero temperature
$P(x)$ is coupled with the threshold distribution $P_c (x)$
by the relationship~\cite{BP2008,BP2010}
\begin{equation}
P(x) = P_c (x) / J_c (x)\,, \;\;\; J_c (x) = \int_{x}^{\infty} d\xi P_c (\xi) \,.
\label{PcP}
\end{equation}
The function $\Gamma (X)$ in Eq.~(\ref{Q5a})
describes the contacts that form again after breaking,
\begin{equation}
\Gamma (X) = \int_{-\infty}^{\infty} d\xi \, P(\xi) \, Q(\xi;X)
\label{Q5aG}
\end{equation}
(the delay time is neglected at this stage),
and $R(x)$ is the (normalized) distribution of stretchings for newborn contacts.
Then, the friction force is given by
\begin{equation}
F(X)=N_c \, k \int_{-\infty}^{\infty} dx \, x \, Q(x;X) \,.
\label{Q6}
\end{equation}

The evolution of the system in the quasi-static limit where inertia
effects are neglected
shows that, in the long term, the initial distribution approaches a
stationary distribution $Q_s (x)$ and
the total force $F$ becomes independent of $X$.
This statement is valid for any distribution $P_c (x)$
except for the singular case of $P_c (x) = \delta (x-x_s)$.

\smallskip
In the present work we concentrate on the steady state (smooth sliding).
In what follows we use $R(x) = \delta(x)$ for simplicity.
The steady-state solution of Eq.~(\ref{Q5a}) is
\begin{equation}
Q (x) = \Theta(x) E_P (x)/C[P] \,,
\label{Q11a}
\end{equation}
where $\Theta(x)$ is the Heaviside step function
($\Theta(x)=1$ for $x \geq 0$ and 0 otherwise),
$E_P (x) = \exp [-U(x)]$, $U(x) = \int_{0}^{x} d\xi \, P(\xi)$, and
$C[P] = \int_{0}^{\infty} dx \, E_P (x)$.
Note also that, in the steady state,
\begin{equation}
\Gamma = 1/C[P] \,,
\label{Gamma0}
\end{equation}
because
$ \int_0^{\infty} d\xi \, P(\xi) E_P (\xi) =
\int_0^{\infty} dU \, e^{-U} =1 $.

The distribution $\widetilde{P}_c (f)$ can be estimated
for the contact of rough surfaces~\cite{GW1966,BP2010}
as well as for the contact of polycrystal substrates~\cite{BP2010,BMnext}:
its general shape may be approximated by the function
\begin{equation}
\label{eq:pcf}
\widetilde{P}_c (f) \propto f^n \exp (-f/f_*) \,,
\end{equation}
where $n \ge 0$ depends on the nature of the interface.
Then, the distribution $P_c(x)$
can be related to the distribution $\widetilde{P}_c (f_s)$
of the static friction force thresholds of the contacts.
If a given contact has an area $A$, then it is characterized by the
static friction threshold $f_s \propto A$ and
the (shear) elastic constant $k \propto \sqrt{A}$
(assuming that the linear size of the contact and its height
are of the same order of magnitude, see~\cite{P1995}
and Appendix A in Ref.~\cite{BP2010}).
The displacement threshold for the given contact is $x_s = f_s /k$,
so that $f_s \propto x_s^2$,
or $df_s /dx_s \propto x_s$.
Then, using 
$P_c (x_s) \, dx_s = \widetilde{P}_c (f_s) \, df_s$,
we obtain
$P_c (x_s) \propto x_s \widetilde{P}_c [f_s (x_s)]$, or
\begin{equation}
P_c (x) \propto x^{1+2n} \exp (-x^2/x_*^2) \,,
\label{PxPf2}
\end{equation}
where $x_*$ may be estimated from experiments
as $N_c k x_* \approx F_s$.
In the SFA/B (surface force apparatus/balance) experiments,
where the sliding surfaces are made of mica,
the interface may be atomically flat over a macroscopic area.
But even in this case the lubricant film cannot be
ideally homogeneous throughout the whole contact area
--- it should be split into domains, e.g., with different orientation,
because this will lower the system free energy due to the increase of entropy.
Domains of different orientations have different values for the
thresholds $f_{si}$,
i.e., they play the same role as asperities in the contact of rough surfaces.

\smallskip
For the normalized distribution of static thresholds given by
Eq.~(\ref{PxPf2}) with $n=1$,
\begin{equation}
P_c (x) = (2/ x_*) \, u^3 e^{-u^2},
\;\;\; {\rm where} \;\;\;
u \equiv x/x_* \,,
\label{nt01}
\end{equation}
we can express the steady-state solution of the master equation
analytically.
In this case
\begin{equation}
J_c (x) = (1+u^2) \, e^{-u^2} \,,
\label{nt02}
\end{equation}
so that at zero temperature we have
\begin{equation}
P(x) = (2/ x_*) \, u^3 /(1+u^2) \,,
\label{nt03}
\end{equation}
\begin{equation}
U(x) = u^2 - \ln (1+u^2) \,,
\label{nt03a}
\end{equation}
\begin{equation}
E_P (x) = J_c (x) = (1+u^2) \, e^{-u^2} \,,
\label{nt03b}
\end{equation}
\begin{equation}
C[P] = x_* /C_0,
\;\;\; {\rm where} \;\;\;
C_0 = \frac{4}{3 \sqrt{\pi}} \approx 0.752,
\label{nt05}
\end{equation}
\begin{equation}
Q(x) = (C_0 / x_*) (1+u^2) \, e^{-u^2} ,
\;\;\; u \geq 0 
\label{nt06}
\end{equation}
and the kinetic friction is
\begin{equation}
f_k \equiv F_k / \left( N_c k \right)
= f_{k0} \equiv C_0 x_* \,.
\label{nt07}
\end{equation}

\smallskip
The ME formalism described above can be extended to take into account various
generalizations of the EQ model, such as temperature effects and
contact aging, which are examined in the following sections.
%

\section{Nonzero temperature}
\label{temperature}

Temperature effects enter in the ME formalism
through their effect on the fraction of contacts that break per
unit displacement of the sliding block, $P(x)$, because thermal
fluctuations allow an activated breaking of any contact
which is still below the
threshold~\cite{S1963,P1995,FKU2004,SW2009,BP2010,BU2010}.
For a sliding at velocity $v$ so that $X=vt$,
the thermally activated jumps can be incorporated in the master equation,
if we use,
instead of the zero-temperature breaking fraction density
$P(x)$,
an expression $P_T (x)$ defined by (see~\cite{BP2010})
\begin{equation}
P_T (x) = {P}(x) + H(x) \,,
\label{temper4}
\end{equation}
where the temperature contribution
is given by
\begin{equation}
H(x)=\frac{\omega}{v} \; e^{k x^2 /2k_B T} \int_{x}^{\infty} d\xi \, P_c (\xi)
\, e^{-k \xi^2 /2k_B T}
\label{temper11}
\end{equation}
for ``soft'' contacts or by
\begin{equation}
H(x)=\frac{\omega}{v} \int_{x}^{\infty} d\xi \, P_c (\xi)
\left( 1-\frac{x}{\xi} \right)^{\frac{1}{2}}
e^{- {k \xi^2 \left( 1- \frac{x}{\xi}
\right)^{\frac{3}{2}} } / {2k_B T} }
\label{temper12}
\end{equation}
in the case of ``stiff'' contacts which have a deep pinning potential
so that their breaking only occurs with a significant probability
when their stretching is close to the threshold.
Here $\omega$ is the attempt frequency of contact breaking,
$\omega \sim 10^{10}$~s$^{-1}$ according to Refs.~\cite{P1995,BU2010}.

For concreteness, in what follows
we assume that the contacts are soft, Eq.~(\ref{temper11}),
and we select $n=1$ in Eq.~(\ref{PxPf2}),
so that $P_c (x)$ is given by Eq.~(\ref{nt01}).


At a nonzero temperature the total rate of contact breaking,
Eq.~(\ref{temper4}),
is equal to $P_T (x) = {P}(x) + (\omega /v) \, h(x)$,
where for the soft contacts
\begin{equation}
h(x) 
= \frac{1+ (1+b) \, u^2}{(1+b)^2} \, e^{-u^2}
\label{nt08}
\end{equation}
with
\begin{equation}
b(T)=\frac{k x_*^2}{2 k_B T} \,.
\label{nt08a}
\end{equation}
%
The condition $b=1$ defines a crossover temperature
\begin{equation}
k_B T_* = \frac{1}{2} \, k x_*^2 \,.
\label{nt09a}
\end{equation}

Then, a straightforward integration gives the function
$U_T (x) = \int_{0}^{x} d\xi \, P_T (\xi) = {U}(x) + \Delta U(x)$, where
\begin{equation}
\Delta U(x) =
S_0 (v,T) \left[ \, {\rm erf} (u) - S_1 (T) \, u \, e^{-u^2} \right] \,
\label{nt09}
\end{equation}
with
\begin{equation}
S_0 (v,T) =
\frac{\omega x_*}{C_0 v} \,
\frac{(1+b/3)}{(1+b)^2} \,
 \label{nt10}
\end{equation}
and
\begin{equation}
S_1 (T) =
\frac{C_0}{2} \, \frac{(1+b)}{(1+b/3)} \,.
\label{nt11}
\end{equation}
The coefficient $S_1 (T)$
weakly changes with temperature
from $S_1 (0) = 2/\sqrt{\pi} \approx 1.128$
to $S_1 (\infty) = C_0 /2 \approx 0.376$.
On the other hand, the coefficient $S_0 (v,T)$
determines whether the effect of temperature is essential or not.
The temperature-induced breaking of contacts is essential
at low driving velocities only, when $S_0 (v,T) \gg 1$.
Thus, the equation $S_0 (v_*, T) = 1$ defines the crossover
velocity:
\begin{equation}
v_* (T) = \frac{\omega x_*}{C_0} \;
\frac{[1+b(T)/3]}{[1+b(T)]^2} \;.
\label{nt12}
\end{equation}
We see that $v_*$ monotonically increases with temperature
as $v_* (T) \approx 0.443 \, \omega x_* T/T_*$ at $T \ll T_*$
and approaches the maximal value $v_* \approx 1.33 \, \omega x_*$
at $T \gg T_*$.

Then, $E_{PT} (x) = e^{-U_T (x)} = (1+u^2) \, e^{-u^2} e^{-\Delta U(x)}$,
and we can find the kinetic friction force:
\begin{equation}
f_k (v,T) = \int_{0}^{\infty} dx \, x E_{PT} (x) \biggr/
\int_{0}^{\infty} dx \, E_{PT} (x) \,.
\label{nt13}
\end{equation}

At a low driving velocity, $v \ll v_*$, we may put
$\Delta U(x) \approx S_2 u$, where
\begin{equation}
S_2 (v,T)=\frac{\omega x_*}{v \, (1+b)^2} \,,
\label{nt14}
\end{equation}
and Eq.~(\ref{nt13}) leads to
\begin{equation}
f_k \approx x_*/S_2 = (v/\omega) (1+b)^2 . 
\label{nt15}
\end{equation}
A linear dependence of the kinetic friction on the driving velocity
at low velocities corresponds to the creep motion due to
temperature activated breaking of contacts
and was predicted in several earlier studies~\cite{S1963,P1995,SW2009,BU2010},
although our approach allowed us to derive it rigorously.
The dependence~(\ref{nt15})
may be interpreted as an effective ``viscosity''
of the confined interface:
\begin{equation}
\eta = \frac{f_k}{v} = \frac{1}{\omega} \left( 1 + \frac{k x_*^2}{2
  k_B T} \right)^2 .
\label{nt15a}
\end{equation}

At a high velocity, $v \gg v_*$, when
$e^{-\Delta U(x)} \approx 1-S_2 u + {1\over2} (S_2 u)^2$,
we obtain
$C \approx 3\sqrt{\pi}/4 -S_2 + (5\sqrt{\pi}/16) S_2^2$
so that
\begin{equation}
f_k \approx f_{k0} \left( 1-C_1 S_2 +C_2 S_2^2 \right) ,
\label{nt16}
\end{equation}
where
$C_1 = 5\sqrt{\pi}/8-C_0\approx 0.356$
and
$C_2 = 16/9\pi -1/3 \approx 0.233$.
Equation~(\ref{nt16}) agrees qualitatively with that found by
Persson~\cite{P1995} in the case of $b \gg 1$.

\begin{figure} 
\includegraphics[clip, width=8cm]{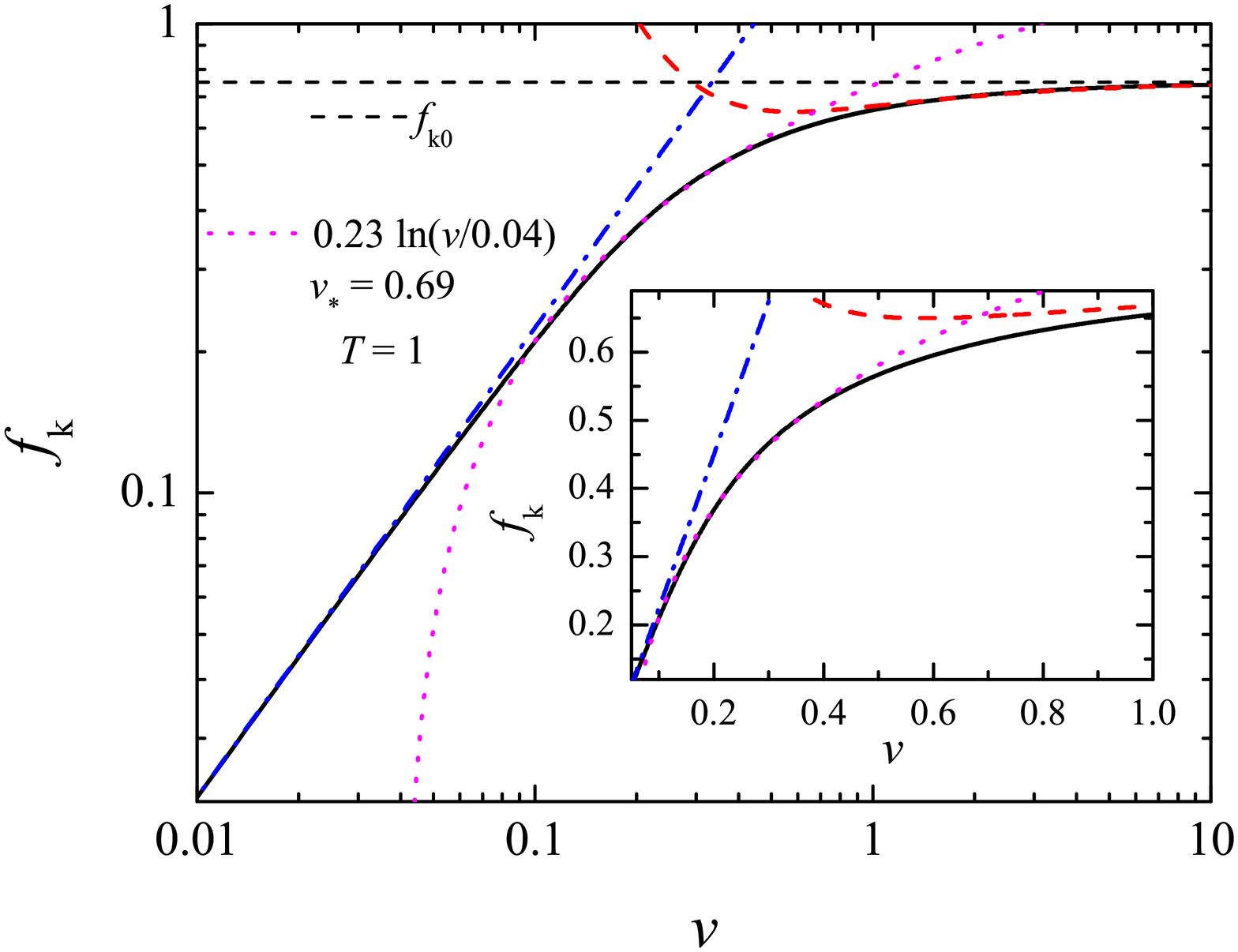}
\caption{\label{A01}(Color online):
The friction force $f_k$ as a function of the driving velocity.
Dash-dotted blue line shows the low-$v$ approximation (\ref{nt15}),
dashed red line shows the high-$v$ approximation (\ref{nt16}),
and dotted magenta line shows a logarithmic fitting.
$k = 1$, $\omega = 1$, $x_* = 1$, $k_B T=1$.
}
\end{figure}
Approximate expressions (\ref{nt15}, \ref{nt16})
together with the numerical integration of Eq.~(\ref{nt13})
are presented in Fig.~\ref{A01}.
Also we showed a logarithmic fitting
which operates in a narrow interval of velocities near the crossover
velocity only.
Persson~\cite{P1995} showed that the logarithmic dependence
may be obtained analytically, only if the $P_c(x)$ distribution
has a sharp cutoff at some $x=x_s$ as, e.g.,
in simplified versions of the EQ model with $P_c(x)=\delta(x-x_s)$.

\begin{figure} 
\includegraphics[clip, width=8cm]{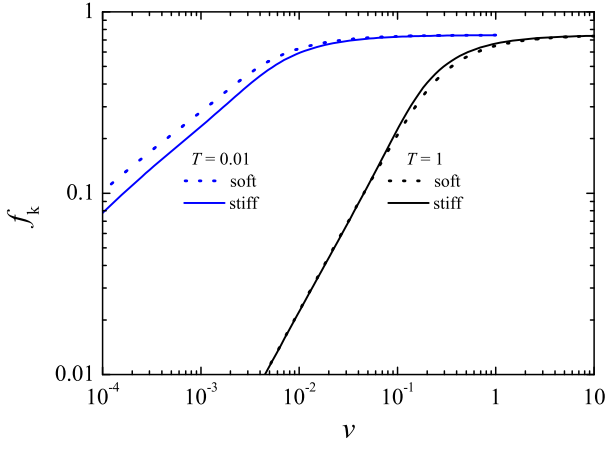}
\caption{\label{A01num}(Color online):
The friction force $f_k$ as a function of the driving velocity $v$
for soft (dotted) and stiff (solid curves) contacts
at low temperature $k_B T=0.01$ (blue)
and high temperature $k_B T=1$ (black).
$k = 1$, $\omega = 1$, $x_* = 1$.
}
\end{figure}
Although we cannot obtain analytical results for the stiff contacts,
we calculated the $f_k (v)$ dependences numerically
(see Fig.~\ref{A01num}),
which shows that the effect remains qualitatively the same.

\section{Aging of contacts}
\label{aging}

The aging of contacts was considered in our work~\cite{BP2010}
where, however, we ignored the temperature-induced breaking of contacts
discussed above in Sec.~\ref{temperature}.
When aging is taken into account
the master equation for $Q(x,X)$ must be completed
by an equation
for the evolution of $P_c(x)$, which in turns affects $P(x)$.
Let the newborn contacts be characterized by a distribution $P_{ci} (x)$,
while at $t \to \infty$, due to aging the distribution $P_c (x)$ approaches
a final distribution $P_{cf} (x)$.
If we assume that the evolution of $P_c (x)$ corresponds to a
stochastic process,
then it should be described by a Smoluchowsky equation
\begin{equation}
\frac{\partial P_c}{\partial t} = D \,  \widehat{L} P_c,
\;\;\; {\rm where} \;\;\;
\widehat{L} \equiv \frac{\partial }{\partial x} \left( B(x) +
\frac{\partial }{\partial x} \right),
\label{L}
\end{equation}
in which the ``diffusion'' parameter $D$ describes the rate of aging,
$B(x)=d\bar{U}(x)/dx$,
and the ``potential'' $\bar{U}(x)$ determines the final distribution,
$P_{cf} (x) \propto \exp \left[ -\bar{U}(x) \right]$,
so that we can write
\begin{equation}
B(x) = -\left[ dP_{cf} (x) /dx \right] /{P_{cf} (x)} \,.
\label{Bx}
\end{equation}
However, because the contacts continuously break and form again when the
substrate moves,
this introduces two extra contributions in the equation determining
$\partial P_c / \partial X$ in addition to the pure aging effect described
by Eq.~(\ref{L}): a term $P(x;X)Q(x;X)$ takes into
account the contacts that break, while their reappearance with the threshold
distribution $P_{ci}(x)$ gives rise to the second extra term in the equation.
Thus, the evolution of $P_c$ is described by the equation
\begin{align}
\frac{\partial P_c (x;X)}{\partial X} - & D_v \widehat{L} P_c (x;X)
+ P(x;X) \, Q(x;X)
\nonumber \\
& =  P_{ci}(x) \, \Gamma (X) \,,
\label{Q15}
\end{align}
where $D_v = D/v$, and $v=dX(t)/dt$ is the driving velocity.
Finally, we come to the set of equations~(\ref{Q5a}--\ref{Q5aG}, \ref{Q15}).
%
For the steady-state regime, Eq.~(\ref{Q15}) reduces to
\begin{equation}
D_v C[P] \, \widehat{L} P_c (x) = P(x) \, E_P (x) - P_{ci}(x)\,,
\label{Q15a}
\end{equation}
where we used Eqs.~(\ref{Q11a}) and~(\ref{Gamma0}).
Taking also into account the identity $ P(x) \, E_P (x) = P_{c}(x)$
\cite{BP2010}, we finally come to the equation
\begin{equation}
D_v C[P] \, \widehat{L} P_c (x) = P_c (x) - P_{ci}(x)\,.
\label{Q15b}
\end{equation}

It was shown \cite{BP2010} that the kinetic friction monotonically
decreases with the driving velocity as
$F_k (v) - F_k (0) \propto - v/D$
in the low-velocity limit and
$F_k (v) - F_k (\infty) \propto D/v$
in the high-velocity case. 
One may expect that at low velocities this decreasing will
compensate the friction increasing due to temperature induced jumps.
The problem, however, is more involved.

When the temperature effects are incorporated,
Eq.~(\ref{Q15b}) for the function $P_c (x)$ in the steady state takes the form
\begin{equation}
D_v \, C[P_T] \, \widehat{L} P_c (x) =
P_c (x) - P_{ci} (x) \,.
\label{star1}
\end{equation}

\begin{figure} 
\includegraphics[clip, width=8cm]{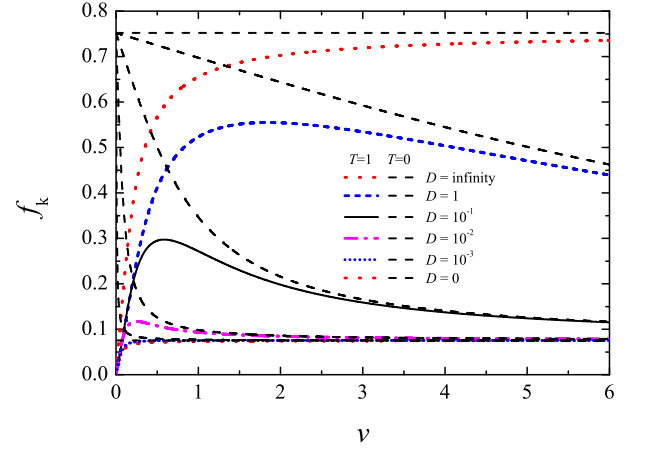}
\caption{\label{A02}(Color online):
The kinetic friction force $f_k$ as a function of the driving velocity $v$
for different values of the aging rate: 
$D = \infty$ (red dotted),
$1$ (blue short-dashed),
$10^{-1}$ (black solid),
$10^{-2}$ (magenta dash-dotted),
$10^{-3}$ (blue short-dotted), and
$D = 0$ (red dotted curve).
$k = 1$, $\omega = 1$, $k_B T=1$;
the initial and final $P_c (x)$ distributions are given by Eq.~(\ref{nt01}) with
$x_{*i} = 0.1$ and $x_{*f} = 1$ correspondingly.
Dashed curves shows the dependences at $T=0$.
}
\end{figure}
\begin{figure} 
\includegraphics[clip, width=8cm]{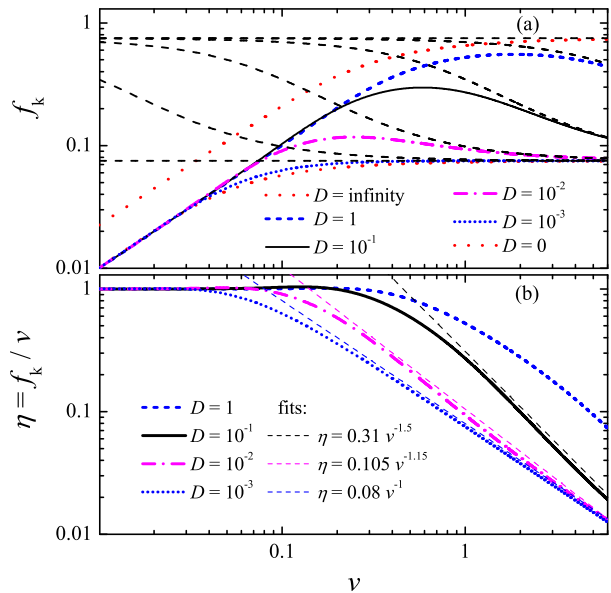}
\caption{\label{A02ab}(Color online):
(a) The same as in Fig.~\ref{A02} in log-log scale.
(b) The dependences of the effective ``viscosity'' $\eta = f_k /v$
on the velocity $v$ (dashed lines show power-law fits).
}
\end{figure}
Numerical solutions of
Eq.~(\ref{star1}) are presented in Figs.~\ref{A02} and~\ref{A02ab}a:
the initial increase of the kinetic friction $F$
with the driving velocity $v$ due to the temperature activated
breaking of contacts 
is followed by the decrease of $F$ due to contacts aging.
Figure~\ref{A02ab}b shows also the dependence of the effective ``viscosity''
$\eta = f_k /v$ on the driving velocity $v$.
It is constant at low velocity and then decreases;
the latter may be approximately fitted by a power law
$\eta (v) \propto v^{-\alpha}$ with the exponent $\alpha$
changing from 1.5 to 1 as $D$ decreases.

\smallskip
Using the definition~(\ref{L}) of the operator $\widehat{L}$
and Eq.~(\ref{Bx}) for the function $B(x)$, the l.h.s.\ of Eq.~(\ref{star1})
may be rewritten as
\begin{equation}
D_v C \, \widehat{L} P_c (x) =
D_v C \, \frac{d}{dx} \left( P_{cf} (x) \,
\frac{d}{dx} \frac{P_{c} (x)}{P_{cf} (x)} \right) \,,
\label{star1a}
\end{equation}
while the r.h.s.\ of Eq.~(\ref{star1}) may be presented as
\begin{equation}
P_c (x) - P_{ci} (x) = - \frac{d}{dx} \left[ J_c (x) - J_{ci}  \right] \,,
\label{star1b}
\end{equation}
where
$J_{ci} (x) = \int_x^{\infty} d\xi \, P_{ci} (\xi)$. 
Using Eqs.~(\ref{star1a}) and (\ref{star1b}), we can find the first
integral of Eq.~(\ref{star1}):
\begin{equation}
D_v \, C[P_T] \, P_{cf} (x) \,
\frac{d}{dx} \left( \frac{P_{c} (x)}{P_{cf} (x)} \right) = J_{ci} (x)
- J_c (x) \,.
\label{star2}
\end{equation}
Integration of Eq.~(\ref{star2}) leads to an integral equation
for the function $P_{c} (x)$:
\begin{equation}
P_{c} (x) = P_{cf} (x) \left\{ 1 + \frac{v}{D C[P_T]}
\int_{0}^x \frac{J_{ci} (\xi) - J_{c} (\xi)}
{P_{cf} (\xi)} \, d\xi \right\} \,.
\label{star2a}
\end{equation}

\smallskip
Substituting
$J_{c} (x) \approx J_{cf} (x) = \int_x^{\infty} d\xi \, P_{cf} (\xi)$
into the r.h.s.\ of Eq.~(\ref{star2a}),
one may  analytically find the low-velocity behavior of the kinetic friction,
for example, the decrease of $f_k$ with $v$ for $T=0$.
At a nonzero temperature, however, aging does not affect
the low-velocity behavior (\ref{nt15}) and only reduces the interval
of velocities where Eq.~(\ref{nt15}) is valid,
as demonstrated in Figs.~\ref{A02} and~\ref{A02ab}.
Indeed, at $v \to 0$ and $T>0$ the main contribution to $P_T (x)$
comes from the function $H(x) \propto \omega /v$ in Eq.~(\ref{temper4}),
which only weakly depends on $P_c (x)$.

\begin{figure} 
\includegraphics[clip, width=8cm]{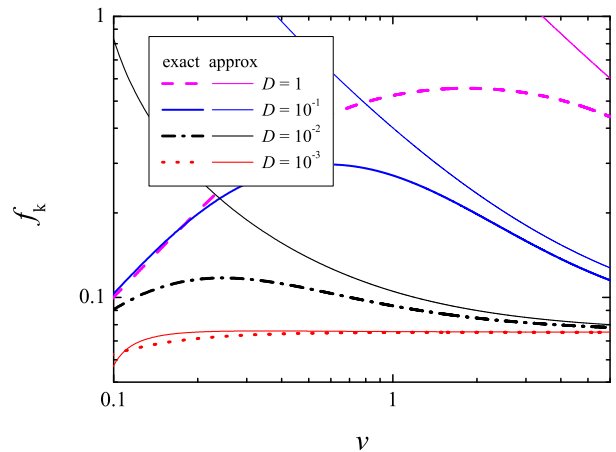}
\caption{\label{A04}(Color online):
The kinetic friction force $f_k$ as a function of the driving velocity $v$
for different values of the aging rate
$D=1$ (magenta dashed),
$10^{-1}$ (blue solid),
$10^{-2}$ (black dash-dotted) and
$10^{-3}$ (red dotted curve)
as compared with approximate expressions (thin solid curves).
$k = 1$, $\omega = 1$, $k_B T=1$;
the initial and final $P_c (x)$ distributions are given by Eq.~(\ref{nt01}) with
$x_{*i} = 0.1$ and $x_{*f} = 1$ correspondingly.
}
\end{figure}
The limit $D_v = D/v \to 0$ may be studied with the help of Eq.~(\ref{star1})
by substituting $P_c (x) \approx P_{ci} (x)$ into its left-hand side.
For the function (\ref{nt01}), this approach gives
\begin{equation}
P_c (x) - P_{ci} (x) \approx -\frac{16 D_v C[P]}{x_{*i} x_{if}^2} \,
u_i e^{-u_i^2} \left( 1- {1\over2} u_i^2 \right)
\label{star3a}
\end{equation}
and
\begin{equation}
J_c (x) - J_{ci} (x) \approx \frac{4 D_v C[P]}{x_{if}^2} \;
u_i^4 e^{-u_i^2} \,,
\label{star3b}
\end{equation}
where $u_i = x/x_{*i}$ and
\begin{equation}
\frac{1}{x_{if}^2} = \frac{1}{x_{*i}^2} - \frac{1}{x_{*f}^2} \,.
\label{star3c}
\end{equation}
Then, taking the corresponding integrals, we obtain
to first order in $v^{-1}$
\begin{equation}
C \approx x_{*i} \left( C_0^{-1} - S_2 + C_{12} S_3 \right)
\label{star4a}
\end{equation}
and
\begin{equation}
f_k \approx \frac{x_{*i}^2}{C} \left[ 1-(C_0 + C_1) S_2 + \widetilde{C}_{12} S_3
\right] \,,
\label{star4b}
\end{equation}
where
\begin{equation}
S_3 (v) = 4 D_v C[P] /x_{if}^2 \,,
\label{star4c}
\end{equation}
$C_{12} \approx 1.324$ and
$\widetilde{C}_{12} \approx 1.789$ are numerical constants.
A comparison of the exact and approximate expressions
is shown in Fig.~\ref{A04}.

\section{Delay in contact formation}
\label{delay}

Finally, let us take into account the delay in contact formation
following the work of Schallamach~\cite{S1963}.
Let $\tau$ be the delay time,
$N$ be the total number of contacts,
$N_c$ be the number of coupled (pinned) 
contacts, and
$N_f = N-N_c$ be the number of detached (sliding) contacts.
The fraction of contacts
that detach per unit displacement of the
sliding block is
$\Gamma (v,T) = \int dx \, P(x) \, Q(x)$,
i.e., when the slider shifts by $\Delta X$,
the number of detached contacts changes by $N_c \Gamma \Delta X$, so that
$N_f = \Gamma v \tau N_c$.
Using $N_c + N_f = N$, we obtain
$N_c = N / (1 + \Gamma v \tau)$
and
$N_f = N \Gamma v \tau / (1 + \Gamma v \tau)$.
If we define $\bar{x} = 1/\Gamma$ and $v_d = \bar{x} /\tau$,
we can write
\begin{equation}
N_c = \frac{N}{1 + v /v_d}
\;\;\;{\rm and}\;\;\;
N_f = \frac{N v /v_d}{1 + v /v_d} \; .
\label{eq:ncnf}
\end{equation}

\begin{figure} 
\includegraphics[clip, width=8cm]{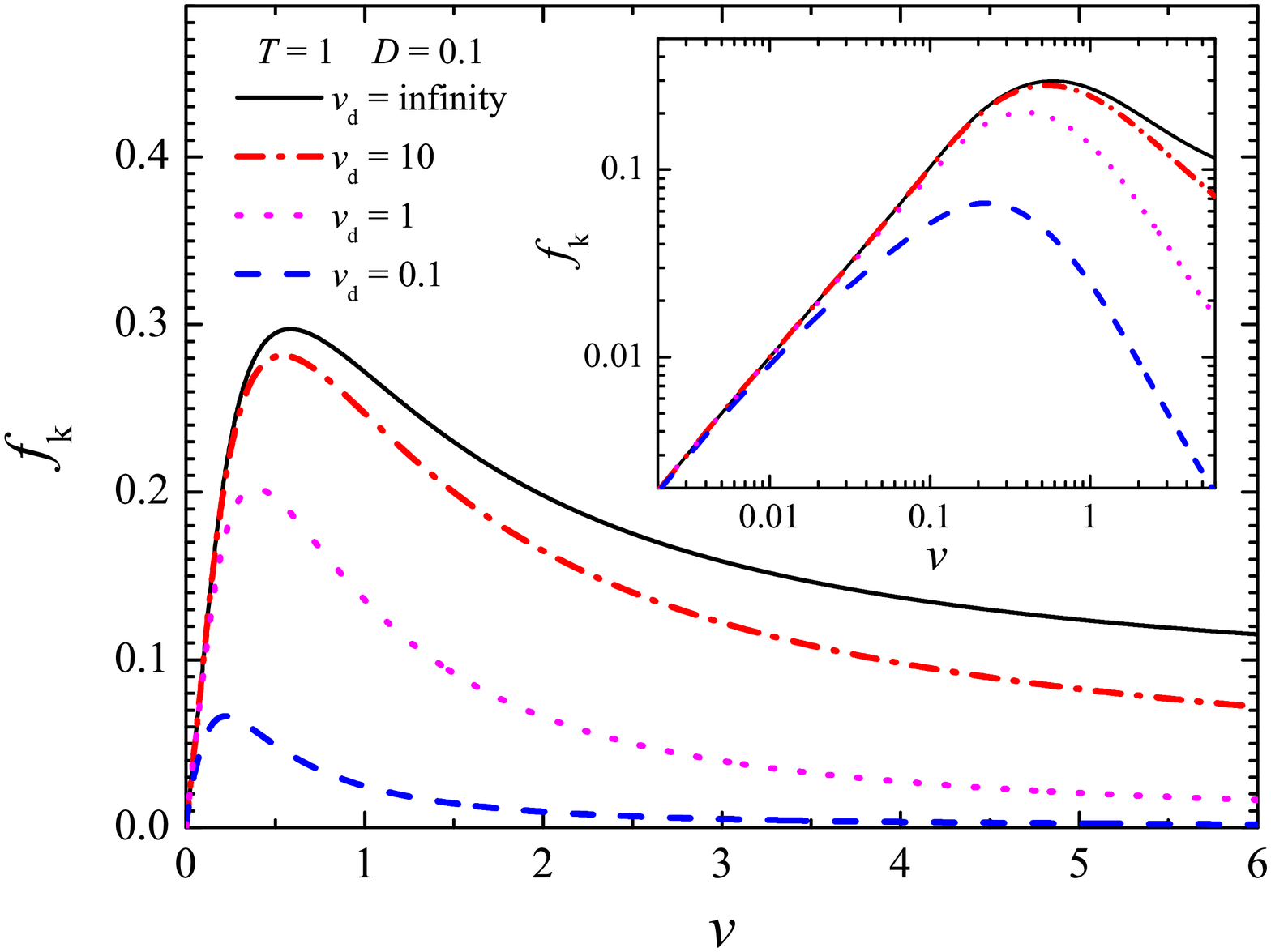}
\caption{\label{A03}(Color online):
The kinetic friction force $f_k$ as a function of the driving velocity $v$
for different values of the delay time:
$v_d = \infty$ (black solid),
$10$ (red dot-dashed),
$1$ (magenta dotted) and
$0.1$ (blue dashed curve)
for $D=0.1$ and $k_B T=1$
(other parameters as in Fig.~\ref{A02}).
Inset shows the same in log-log scale.
}
\end{figure}
The coupled contacts produce the force $f_k$
defined above by the steady-state solution of
the master equation.
The combined dependence which incorporates
temperature effects, aging and delay in contact formation,
is shown in Fig.~\ref{A03} for different values of the parameter $v_d$.

However, above we assumed that
the sliding contacts experience zero friction,
while these contacts may experience
a viscous friction force
$f_l = \eta_l v$, where $\eta_l$ corresponds to the (bulk) viscosity
of the liquid lubricant.
In this case the kinetic friction should be additionally multiplied
by a factor $\beta (v) = 1 + v^2 /v_{\eta} v_d$,
where $v_{\eta} = f_k / \eta_l$ ($v_{\eta} \gg v_d$).
Such a correction may be expected at huge velocities only, e.g.,
for $v \sim 1$~m/s.
In this case the function $F_k (v)$, after decreasing,
reaches a minimum at a velocity
$v_0 \approx (v_{\eta} v_d)^{1/2}$,
and then increases according to a law $f_k (v) \propto \eta_l v$.
Note that the viscous friction which comes from the excitation of phonons
in the substrates, as shown in MD simulation \cite{BN2006},
may also depend on the velocity, e.g., as $\eta_l \propto v^4$.

\section{Making the link with experiments}
\label{experiments}

For a real system, the results presented in the previous sections
allow the calculation of the kinetic friction force $F_k (v,T)$
provided the parameters of the model are known. In this section
section we examine how they can be evaluated from
experiments.

The contact parameters $k$ and $\omega$
may be estimated with the help of elastic theory~\cite{LL1970}.
Let us assume that a contact has a cylinder shape
of height $h$ (the thickness of the interface) and radius $r_c$,
so that it is characterized by the section $A_i=\pi r_c^2$,
the (geometrical) inertial momentum $I=\pi r_c^4 /4$,
a mass density $\rho$ and a Young modulus $E$.
If the cylinder foot is fixed and a force $\Delta f$ is applied to its top,
the latter will be shifted on the distance $\Delta x = \Delta f h^3 /3EI$
(
the 
problem of bending pivot, see Sec.~20, example~3
in Ref.~\cite{LL1970}
).
Thus, the effective elastic constant of the contact is $k=\Delta
f/\Delta x=3EI/h^3$.
The minimal frequency of bending vibration of the pivot with one fixed end
and one free end, is given by
$\omega \approx (3.52/h^2) (EI/ \rho A_i)^{1/2}$
(see Sec.~25, example~6 in Ref.~\cite{LL1970}).

Next, let $a$ be the average distance between the contacts,
so that the total area of the interface is $A=Na^2$,
and introduce the dimensionless parameter $\gamma_c = r_c /a$
($\gamma_c < 0.5$).
The threshold distance $x_*$ may be estimated as follows.
At the beginning, when all contacts are in the unstressed state,
the maximal force the slider may sustain is equal to
$F_* \approx Nkx_*$
(this force corresponds to the first large stick spike in the $F(t)$
dependence at the beginning of stick-slip motion at low driving).
Thus, we obtain that $kx_* \approx a^2 \sigma_*$, where
$\sigma_* = F_* /A$ is the maximal shear stress.

\smallskip
Let us consider a contact of two rough surfaces
and assume that $a=h=r_c$.
Then we obtain
\begin{equation}
\omega \approx \frac{1.76 \sqrt{E/\rho}}{r_c}
\label{eq:omeg}
\end{equation}
for the attempt frequency,
\begin{equation}
k = (3\pi /4)Er_c
\end{equation}
for the contact elasticity, and
\begin{equation}
x_* = r_c^2 \sigma_* /k
\end{equation}
for the threshold distance.
For steel substrates we may take
$\rho = 10^4$~kg/m$^3$ for the mass density,
$E=2 \times 10^{11}$~N/m$^2$ for the Young modulus,
and $\sigma_c = 10^9$~N/m$^2$ for the plasticity threshold.
Assuming that
$\sigma_* = \sigma_c$ and $r_c \approx 1$~$\mu$m,
we find that
$\omega \approx 7.9 \times 10^9$~s$^{-1}$,
$k \approx 4.7 \times 10^5$~N/m,
$x_* \approx 2.1 \times 10^{-9}$~m,
$b \approx 2.7 \times 10^{8}$ for room temperature
(i.e., $b \gg 1$), so that the crossover velocity is quite low,
$v_* \approx 0.03$~$\mu$m/s.

\smallskip
Now let us consider a lubricated system, e.g., the one
with a few OMCTS layers as studied by Klein~\cite{K2007} and
Bureau~\cite{B2010},
and assume that the lubricant consists of solidified islands which
melt under stress
as proposed by Persson~\cite{P1993b}.
In this case, instead of using the Young modulus,
let us assume that $x_* = r_c$;
this allows us to find the parameter $EI = ah^3 \sigma_* /3\gamma_c$.
Then, the elastic constant is
$k = a \sigma_* /\gamma_c$,
the attempt frequency is
\begin{equation}
\omega \approx 1.15 \, \sqrt{\sigma_* /ah \rho \gamma_c^3} \,,
\end{equation}
the parameter $b$ is given by
\begin{equation}
b \approx \gamma_c a^3 \sigma_* / 2k_B T \,,
\end{equation}
and in the case of $b \gg 1$ the crossover velocity is
\begin{equation}
v_* \approx \gamma_c \omega a /3C_0 b
\approx k_B T / \sqrt{a^5 h \rho \sigma_* \gamma_c^3} \,.
\end{equation}

For a four-layer OMCTS film~\cite{K2007} one may take
$\rho = 956$~kg/m$^3$,
$h \approx 3.5 \times 10^{-9}$~m,
$F_* \approx 2 \times 10^{-5}$~N and $A \approx 10^{-10}$~m$^2$ so that
$\sigma_* \approx 2 \times 10^{5}$~Pa.
Assuming
$\gamma_c = 0.5$ and $a \approx 1$~$\mu$m,
we obtain for room temperature,
$k_B T = 4 \times 10^{-21}$~J, that
$\omega \approx 8 \times 10^8$~s$^{-1}$ and
$b \approx 1.25 \times 10^7$,
i.e.\ this system is in the low-temperature limit too,
although the crossover velocity is much higher than for rough surfaces,
$v_* \approx 16$~$\mu$m/s.

\begin{figure} 
\includegraphics[clip, width=8cm]{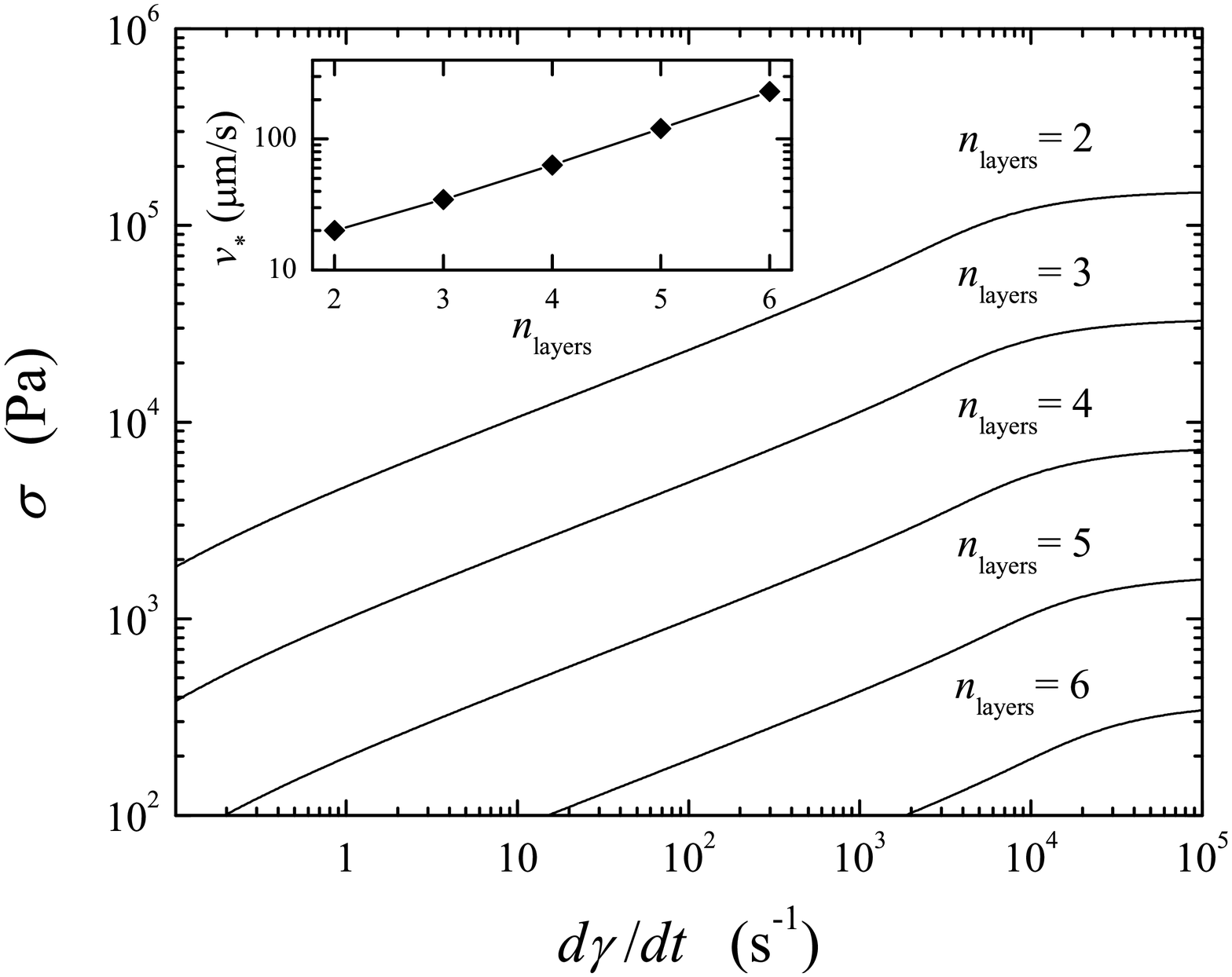}
\caption{\label{A05}
The shear stress $\sigma$ as a function of the shear rate $\dot{\gamma}$
for different values of the OMCTS film thickness
from $n_{\rm l}=2$ to 6 monolayers.
Inset: the crossover velocity $v_*$ as a function of the number of layers.
}
\end{figure}
Moreover, we may calculate the dependence $f_k (v)$
for different thicknesses of the lubricant film.
If the film consists of $n_{\rm l}$ layers,
then the film thickness is $h=n_{\rm l} d$,
where $d \approx 8.75$~\AA %
~is the diameter of the OMCTS molecule.
Let us assume that the maximal shear stress exponentially decreases
with the number of layers according to the results of
MD simulation~\cite{BN2006},
$\sigma_* = \sigma_0 e^{- \beta n_{\rm l}}$,
where $\beta \sim 1$ is a numerical constant.
Taking $\sigma_0 = 4 \times 10^6$~N/m$^2$ and $\beta = 1.5$,
we obtain the dependences of the shear stress
$\sigma = F_k /A$ on the shear rate
$\dot{\gamma} = v/h$ shown in Fig.~\ref{A05},
which may be compared with the experimental dependences (Fig.~2a)
of Bureau~\cite{B2010}.

Note that our approach may overestimate the value of
the crossover velocity $v_*$.
First, the crossover will occur earlier
if the delay and/or aging effects play a significant role.
Besides, at low temperatures the stiff contacts lead to higher ``viscosity''
and lower values of $v_*$ than the soft contacts considered above
(see Fig.~\ref{A01num}).
Second, we completely ignored the elastic interaction between the contacts.
If the latter would be incorporated, a breaking of one contact
may stimulate neighboring contacts to break as well, i.e.,
the value of the parameter $a$ should describe such a cooperative
``contact'' size
which may be much larger than those of individual ones.

\smallskip
Giving a quantitative evaluation of the influence of aging on the
velocity dependence of the friction coefficient is harder than for the
temperature dependence due to insufficient experimental data. Aging
appears to cause a decrease of friction as velocity
increases, and thus, when such a behavior is observed
experimentally~\cite{CRPS2006,Schirmeisen} it can be considered as a
strong indication of the presence of aging. Our analysis indicates
that the combined effect of temperature and aging leads to a maximum
in the friction coefficient versus velocity. Therefore, when aging is
manifested by a decreasing friction versus velocity, extending the
experiments to lower velocities and temperatures might detect the
maximum and thus provide some quantitative data to evaluate the aging
parameters.

\smallskip

Although the aim of our work was to find the dependence
of the kinetic friction on the driving velocity,
our approach allows us to find the dependence on temperature as well.
However, the behavior of a real tribological system is more involved,
because all parameters may
depend on temperature $T$ in a general case.
For example,
the delay time $\tau$ may exponentially depend on $T$ if the formation
of a new contact is an activated process~\cite{S1963};
the same may be true for the aging rate $D$.
In this case one may obtain
a nonmonotonic temperature dependence of friction with, e.g.,
a peak at cryogenic temperatures~\cite{BU2010}.

\section{Conclusion}
\label{concl}

In this study we determined the dependence of the kinetic friction force
in the smooth sliding regime on the driving velocity.
In a general case, the friction linearly increases with the velocity
(this creep motion may be interpreted as an effective ``viscosity''
of the confined film),
passes through a maximum
and then decreases due to delay/aging effects. The decay may be
followed by a new growth in friction in the case of
liquid lubricant.
Estimation showed that for the contact of rough surfaces,
the initial growth of friction should occur at quite low velocities,
$v \ll 0.1$~$\mu$m/s,
so that for typical velocities the friction is independent on velocity
in agreement with the Coulomb law.
However, for the case of lubricated friction with a thin lubricant film
which solidifies due to compression, the $f_k (v)$ dependence
is essential, and the linear dependence may stay valid up to velocities
$v \sim 10 \div 10^3$~$\mu$m/s.
At higher velocities the growth saturates
and the $f_k (v)$ dependence may be fitted by a logarithmic law.
The latter velocity interval is narrow if the distribution
of static thresholds is wide;
the logarithmic law may be found analytically for a wide interval of velocities
when the thresholds are approximately identical, i.e., for the
singular distribution
$P_c (x) = \delta (x-x_s)$.

We emphasize that our approach is  only valid for a system with many contacts,
for example, $N > 20$ at least~\cite{BU2010}.
When the contact is due to a single atom as it may occur
in the AFM/FFM devices, the friction can be accurately described by the
Prandtl-Tomlinson model and should follow the logarithmic $f_k (v)$ dependence,
$f_k (v) \propto ( \ln v/v_0 )^{2/3}$~\cite{JHFS2010}.
But if the AFM/FFM tip is not too sharp so that the contact is due to
more than one atom,
the logarithmic dependence is only approximate and, moreover,
for some systems the friction may decrease with the velocity
which has to be attributed to the aging/delay effects
\cite{RGBMB2003,CRPS2006,HAHSS2007}.

In this work we had in mind that contacts correspond to real asperities
in the case of the contact of rough surfaces
or to ``solid islands'' for the lubricated interface.
However, the ME approach also operates when the contact is due to
long molecules which are attached by their ends to both substrates.
Such a system was first studied by Schallamach~\cite{S1963}
and then further investigated by Filippov \textit{et al.}~\cite{FKU2004},
Srinivasan and Walcott~\cite{SW2009}, and
Barel \textit{et al.}~\cite{BU2010}.
Note that when all molecules are identical,
they are characterized by the same static threshold, i.e.,
this system is close to the singular one,
where the logarithmic $f_k (v)$ dependence has to have a wide interval
of operation.

\smallskip
Finally, let us discuss restrictions of our approach.
First of all, we assumed the somehow idealized case of wearless friction;
wearing may mask the predicted dependences.
Besides, the interface is heated during sliding;
this effect is hard to describe analytically as well as to control
experimentally.
Then, we did not estimated the delay/aging parameters;
moreover, these parameters, e.g., the delay time $\tau$,
may depend on the driving velocity $v$.
Besides, we assumed the simplest mechanism of aging
described by the Smoluchowsky equation,
while the real situation may be more involved, e.g.,
it may correspond to the Lifshitz-Sl\"ozov mechanism~\cite{BP2010}.
Also, we assumed that the reformed contacts appear in the unstressed state,
$R(x) = \delta (x)$ in Eq.~(\ref{Q5a}),
which may not be the case in real systems.

The most important issue, however, is the incorporation
of the elastic interaction between the contacts
as well as elastic deformation of substrates at sliding.
This point certainly deserves a detailed investigation
and is the topic of our future work.

\acknowledgments
This work was supported by CNRS-Ukraine PICS grant No.~5421.


\end{document}